\documentclass[sigconf]{acmart}
\usepackage{amsmath}
\usepackage{amssymb}
\usepackage{color}
\usepackage{subfigure}
\usepackage{multirow}

\newcommand{\vX}{\mathbf{X}}
\newcommand{\vY}{\mathbf{Y}}

\newcommand{\vD}{\mathbf{D}}
\AtBeginDocument{%
  \providecommand\BibTeX{{%
    \normalfont B\kern-0.5em{\scshape i\kern-0.25em b}\kern-0.8em\TeX}}}

\copyrightyear{2020}
\acmYear{2020}
\setcopyright{acmcopyright}\acmConference[MM '20]{Proceedings of the 28th ACM International Conference on Multimedia}{October 12--16, 2020}{Seattle, WA, USA}
\acmBooktitle{Proceedings of the 28th ACM International Conference on Multimedia (MM '20), October 12--16, 2020, Seattle, WA, USA}
\acmPrice{15.00}
\acmDOI{10.1145/3394171.3413921}
\acmISBN{978-1-4503-7988-5/20/10}

\begin{document}
\fancyhead{}
\title{A Human-Computer Duet System for Music Performance}

%\email{webmaster@marysville-ohio.com}
%\affiliation{%
%  \institution{Institute for Clarity in Documentation}
%  \streetaddress{P.O. Box 1212}

%\author{Lars Th{\o}rv{\"a}ld}
%\affiliation{%
%  \institution{The Th{\o}rv{\"a}ld Group}
%  \streetaddress{1 Th{\o}rv{\"a}ld Circle}
%\email{larst@affiliation.org}

%\affiliation{%
%  \institution{Inria Paris-Rocquencourt}

%\affiliation{%
% \institution{Rajiv Gandhi University}

\author{$^\mathsection$Yuen-Jen Lin, $^\mathsection$Hsuan-Kai Kao, $^\mathsection$Yih-Chih Tseng, $^\dagger$Ming Tsai, $^\mathsection$Li Su}
\affiliation{%
  \institution{$^\mathsection$Institute of Information Science,* Academia Sinica, Taipei, Taiwan,\authornote{This work is supported by the Automatic Music Concert Animation (AMCA) project of the Institute.} $^\dagger$KoKo Lab, Taipei, Taiwan}
  \streetaddress{Anonymous Address}}
\email{
r05921045@ntu.edu.tw; hsuankai@iis.sinica.edu.tw; b99901190@gmail.com; cater.t@kokolab.com.tw; lisu@iis.sinica.edu.tw}
%  \city{Anonymous City}

%\affiliation{%
%  \institution{Palmer Research Laboratories}
%  \streetaddress{8600 Datapoint Drive}

%\renewcommand{\shortauthors}{Anonymous authors}

\begin{abstract}
Virtual musicians have become a remarkable phenomenon in the contemporary multimedia arts. However, most of the virtual musicians nowadays have not been endowed with abilities to create their own behaviors, or to perform music with human musicians. In this paper, we firstly create a virtual violinist, who can collaborate with a human pianist to perform chamber music automatically without any intervention. The system incorporates the techniques from various fields, including real-time music tracking, pose estimation, and body movement generation.
In our system, the virtual musician's behavior is generated based on the given music audio alone, and such a system results in a low-cost, efficient and scalable way to produce human and virtual musicians' co-performance. The proposed system has been validated in public concerts. Objective quality assessment approaches and possible ways to systematically improve the system are also discussed. %to evaluate how precise the virtual musician acts in performing music with humans.
\end{abstract}
%%to create all the virtual musician's behaviors, which are generated only from the music to be performed. Such a 


\begin{teaserfigure}
\centering
  \includegraphics[width=0.9\textwidth]{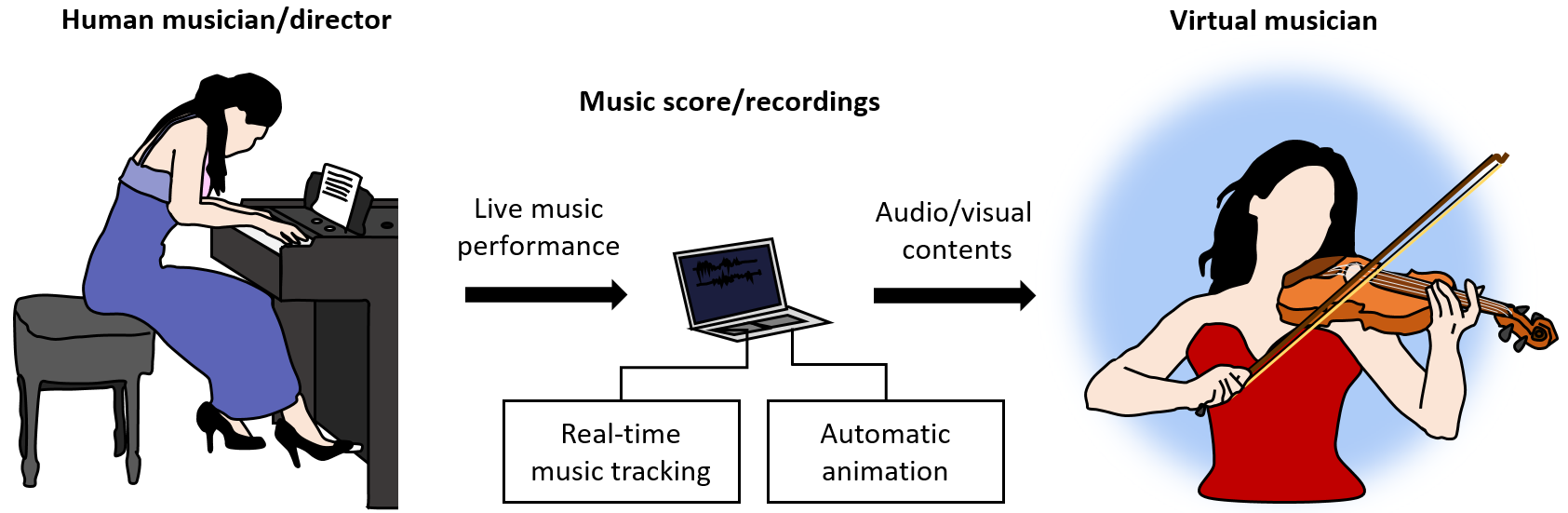}
  \caption{Illustration of the automatic concert animation system.}
  \Description{Enjoying the baseball game from the third-base
  seats. Ichiro Suzuki preparing to bat.}
  \label{fig:teaser}
\end{teaserfigure}

\maketitle

\section{Introduction}

Let's imagine %Consider 
a music concert, where a human musician sits on the stage and performs a duet together with a virtual musician %appearing 
on the screen. The virtual musician can move its body by itself. Its right hand takes a bow to play a virtual violin. Its right hand moves up and down in a way similar to that a real violinist plays the music piece. The virtual musician also follows the human musician's tempo, and make the two voices of the duet to be synchronized and harmonized. All these behaviors of the virtual musician are automatically generated or triggered simply from the music: once the concert program is determined, the human musician can practice, rehearse, and perform music with the virtual musician like with a real human, by simply following the music content (see Figure \ref{fig:teaser}). 

The concept of virtual musician, or more broadly speaking, virtual human character, has become increasingly popular in the past few years. Virtual musicians and related performance types have unlocked great potential in social media and interactive multimedia arts. Virtual idols such as Hatsune Miku \cite{guga2014virtual}, Luo Tianyi \cite{yin2018vocaloid}, and thousands of on-line VTubers\footnote{\url{https://www.bbc.com/worklife/article/20181002-the-virtual-vloggers-taking-over-youtube}} play music either on live streaming platforms or in real-world venues, sometimes even interacting with human musicians \cite{shirai2019reality}. %\footnote{See \url{https://youtu.be/AeAyGchd2LE} for an example. Performance starts at 5'20".} 
Some of the performance videos have even earn them millions of view counts.\footnote{See \url{https://youtu.be/S8dmq5YIUoc} for an example.} As the venues and audiences of virtual music performance are both scaling up, it is expected that a low-cost, personalized, and automatic system which can animate virtual musicians' performance according to the music content, and enforce the interaction between human musicians and virtual ones will be in urgent need. 

For most of the personalized \emph{performance animation} tools such as the Character Animator CC\footnote{\url{https://www.adobe.com/products/character-animator.html}} and many others on making virtual characters, real-time motion capture is employed to capture a human character's facial and body movement data, and rigging technique is then used to convert such data into the virtual character. Being efficient for small-scale production, such a workflow is however limited in producing the content that scales, for example, music performance. As Figure \ref{fig:teaser} shows, a music performance usually incorporates two or more instrument tracks. Recruiting human musicians specialized on various instruments to perform music pieces for every new production is tedious and inefficient. Building a full-interaction mechanism between human and virtual musicians in during music performance is also a less noticed topic; in most of the human-computer collaborative performance, the music is performed simply with a fixed tempo, i.e., in karaoke mode \cite{goto1996jazz}.

Recently, performance animation without the aid of human characters has caught attention. Animation can be generated from texts, voice, and music. For example, %Emerging 
music-to-body-movement generation techniques are expected to solve the difficulty in obtaining professional musicians' motion data: this %. More specifically, music-to-body-movement 
technique aims at mapping a music recording (usually a solo with known instrument type) into a skeleton sequence, and this sequence represents a reasonable movement of a human musician in playing the same music piece with the same instrument. Recently developed motion generation models have shown great potential in generating the movements of violinists \cite{shlizerman2018audio, liu2020body} and pianists \cite{shlizerman2018audio, li2018skeleton}. With such techniques, motion data for animating the virtual musician can be generated directly from the music content, without the need to have human musicians to record the motion data. 

Enforcing the interaction between human and virtual characters during music performance is needed not only in animation. Rather, it has been widely discussed in music information retrieval, robotics, new interfaces for music expression, and multimedia arts. For example, score following and automatic accompaniment systems utilize real-time audio tracking algorithms to make a computer perform accompaniment following the human musician's expressive tempo \cite{arzt2010simple,arzt2010towards}. Alternatively, human musicians also use their body movements, sounds, or physiological signals to trigger, control, and manipulate the virtual musician's sound or actions. \cite{mizumoto2010integration,nagashima2003bio,miranda2005toward,yuksel2015braahms}. Providing high flexibility for artists to design the audio, visual, and interactive effects in the performance, the latter approach however requires the installation of various kinds of sensor devices and data collection systems, and is less suitable for low-cost production such as the performance of an individual VTuber. %Besides, it is still challenging for automatic accompaniment to process audio in real-world environment.

%. Such interaction mechanism is typically built upon real-time and accurate recognition of human musicians' behaviors. Multi-sensory data collection systems can help achieve this requirement and provide more flexibility in rendering the audio or visual effects in the performance. The behaviors of the virtual musician are triggered passively, for example, virtual musicians' body movements are either pre-defined by choreographers during the production process, or designed to responses to a set of stimuli. However, which are less capable to express high-level music semantics such as structure and emotion []. 

In this paper, we focus on the incorporation of automatic animation and real-time music tracking techniques into a virtual musician system. In our proposed system, the human musician and virtual musician communicate with each other by means of music itself, without intervention from manual control or sensor data. Particularly, we emphasize the following three contributions: 1) the virtual musician's behaviors are automatically created from music, 2) the only data required to trigger the interactive music performance is the music content itself, and 3) evaluation metrics for quality assessment on interaction and generation results can be given for further development. We focus on the scenario of \emph{human-computer duet}, as shown in Figure \ref{fig:teaser}, where a human pianist plays duet with a virtual violinist. %The proposed system can operate on standard laptops.
To the best of our knowledge, this work represents the first attempt to bridge the technical gaps among automatic accompaniment, music-to-body-movement generation, and animation. 
%Thanks to the advance of generation models with deep learning in recent years [], we introduce the technique of body movement generation from music signals [], an emerging technique allowing one to generate 3-D body skeletons that fits the music semantics for animating a virtual musician. Also, with the development of music information retrieval (MIR), an on-line audio-to-score alignment (i.e., music tracking) technique \cite{dixon2005live} is also employed to allow the human musician to play music in natural speed while keeping all the actions of the virtual musician synchronized. 
%Such a virtual musician generation system could reduce the cost of human choreography and multi-sensory devices, and is efficient in creating music programs for the vast amounts of VTuber producers with its light-weighting and portable characteristics. 
 
The rest of this paper is organized as follows. Section \ref{sec:related} gives a review on the techniques related to human-computer duet, automatic animation, and interactive multimedia arts. Section \ref{sec:overview} describes the whole system. Technical details of real-time music tracking and body movement generation are described in Section \ref{sec:music_tracker} and Section \ref{sec:body}, respectively. After that, Section \ref{sec:integration} deals with the integration part, including model binding and performance rendering. Real-world demo records and quality assessment of the system are reported in Section \ref{sec:result}. Finally, we conclude our work in Section \ref{sec:conclusion}. 
 
%To specify the main contributions in this paper, we focus on our development of the body movement generation technique (Section X), the on-line music tracking technique (Section X), and the integration of the whole system (Section X). An official demonstration of this system and experimental studies will also be reported (Section X). Possible extensions and the potential of this system in future multimedia industry will also be discussed in Section X. 

%[Most of these studies consider human-computer performance as a recognition problem, rather than the generation problem.]

%, are further gaining their influence in the pop culture. For example, the popular virtual idol Hatsune Miku earns millions of views for each of her YouTube released video. According to the statistics, there are thousands of VTubers on live for every day. One of the most famous VTuber, got XXX follows by the date of 2020. These virtual idols and VTubers perform music on screen or on a real-world venue, and they usually interact with audiences or even real human musicians. 
\section{Related work}
\label{sec:related}

%There is no unique definition for a virtual musician, one could agree that a virtual musician needs to simulate musicians' behaviors, or interact with human. 
Our review of the related work mainly focuses on real-time music tracking and automatic animation, the two critical techniques to characterize the proposed virtual musician system. %Other topics such as music generation and multi-modal interaction are also summarized although they are not in the scope of this paper.

%In this paper, we consider the scenario that the music being performed is already known, and both human and machine are informed with the scores being played. Music generation is therefore out of the scope of discussion and will be left as future work. 

\subsection{Real-time music tracking} 

There are three ways to synchronize human's and machine's performances on live. The first way is letting human follow machine's tempo (usually a constant tempo), and the second way is letting machine follow human's tempo (usually an expressive tempo varying with time and with performance). The first two ways can both be combined with the third one, that utilizes specific gestures, sounds or sensor signals as flag events to re-synchronize the performance \cite{goto1996jazz,taki2000real}. The first way guarantees a stable performance, while the second way, usually known as the \emph{automatic accompaniment} technique, provides great flexibility for human's interpretation of music during performance. Since its first appearance in the 1980s \cite{dannenberg1984line}, automatic accompaniment has been widely studied in various scenarios. Methods for music tracking include state-space models such as Hidden Markov Models (HMM) \cite{cont2006realtime}, Monte Carlo inference \cite{montecchio2011unified}, and particle filters \cite{otsuka2011real}; dynamic time warping (DTW) models such as online DTW (ODTW) \cite{dixon2005live,arzt2010simple,arzt2010towards}, windowed DTW (WTW) \
\cite{macrae2010accurate}; and parallel DTW (PDTW) \cite{rodriguez2016tempo,alonso2017parallel,wei2018online}, and also reinforcement learning approaches \cite{dorfer2018learning}. It should be noted that real-time music tracking has been well-developed technology on MIDI data, but is still highly challenging on audio data an on various classes musical instruments, as the tracking results are sensitive to the variation of audio features. %This is because  as humans are sensitive to any mismatch among the performance contents, even if the mismatch is in only a few milliseconds []. 
Therefore, developed systems and products for audio-based real-time music tracking are still rarely seen, except for some examples such as \emph{Antescofo}'s \emph{Metronaut},\footnote{\url{https://www.antescofo.com/}} which is targeted for automatic accompaniment in music practice.

A number of metrics have been proposed to evaluate a real-time music tracking system. These metrics are mostly based on measuring the latency/ error of every note event \cite{orio2003score, cont2007evaluation,otsuka2010design}, or calculating the number of missing/ misaligned note during the process of score following \cite{cont2007evaluation,arzt2010simple,arzt2012adaptive}. There are, however two major issues in such evaluation methods. First, the performance of score following cannot fully represent the performance of a automatic accompaniment system operating in real-world environments, as it ignores the latency introduced in sound synthesis, data communication, and even reverberation of the environment. Second, note-level evaluation is suitable only for hard-onset instruments such as piano, while it is limited for soft-onset instruments such as violin, as the uncertainty of violin onset detection could propagate errors in the final evaluation results. To solve these issues, we firstly propose an experimental setup which allows evaluation of the system in a real-world environment. Further, we provide frame-level evaluation approach for general types of instrument, with intuitive visual diagrams that demonstrate how the system interacts with human during the performance.
%Limited systematic method have been proposed in previous work to evaluate the robustness and preciseness of the score following algorithm \cite{orio2003score, cont2007evaluation}. In \cite{cont2007evaluation}, in order to calculate the assessment metrics, five basic event measures were defined. Unlike the above-mentioned method, our approach does not focus on finding the respective onset times of the live performance. Mainly, the current state-of-the-art onset detection method accuracy is still unsatisfying; conceivably, violin having a relatively slow attack time will present erratic results in extracting the note events. Instead, we provide not only quantitative measures by using frame-wise approach, but also intuitive visual diagram that demonstrates how our system performs as a function of time. We describe our approach in detail in Section 7.

%The idea to let human collaborate with machine to play music has a history of more than 30 years. 
%In 1984, Roger B. Dannerberg proposed the first automatic accompaniment system. Since then, automatic accompaniment has been widely investigated, though few of them can be applied in the real-world scenario, as the unstability that deteroriate the art of time is almost unavoidable.

%As above, researches on automatic accompaniment typically consider the scenario that a computer and a human musician play duets together. 
In the following of this paper, we use the term \emph{human-computer duet} or \emph{real-time music tracking} when referring to the automatic accompaniment technique. The reason we avoid using the term ``auto accompaniment'' is that in our scenario, the machine's (virtual musician's) role is not limited to accompaniment; the virtual musician can play either accompaniment or main melody.

%because it usually implies a ``master-slave relationship:'' the role of the machine is playing the accompaniment rather than the main melody part. In our work, the main melody is performed by the computer. 
\vspace{-0.3cm}
\subsection{Automatic animation}

%Automatic animation tools such as is used for a wide range of animation scenarios (by means of motion capture). Automatic performance animation tools such as the Adobe Character Animator CC\footnote{\url{https://www.adobe.com/products/character-animator.html}} have changed the workflow of animation. Creating an animated character to mimic a specific human performance. 
Performance animation tools employ motion capture and model rigging technology to create animated characters that mimics specific human performances \cite{chai2005performance,starck2007surface,xia2017survey,limbu2018using}. 
New developments that support multi-modal input such as texts, voice, and music to automate the workflow of performance animation have also received wide attention. For text-to-animation tasks, linguistic rules and action representations are derived from screenplays and then are used to generate pre-visualization in animation \cite{zhang2019generating}. The recently proposed \emph{TakeToons} system achieves efficient animation by incorporating motion capture with a story model, the latter is learned from the scripts with annotation of relevant events such as character actions, camera positions, and scene backgrounds \cite{subramonyam2018taketoons}. 
For video-to-animation tasks, a recent work further applies pose estimation and clustering technique to convert a video into 2-D animation \cite{willett2020pose2pose}. 
For audio-to-animation tasks, animation is synthesized based on mimicking onomatopoeia sounds with audio event recognition techniques \cite{nivaggioli2019animation}. 
The generative adversarial network (GAN) is also applied to convert conversational speech signals to gestures \cite{ginosar2019learning}.
%-	From video (actors’ performance) and video-derived contents (limb, hand, …) -> Pose estimation, pose generation
%-	From audio (speech, music) -> audio-to-gesture
%-	Generating visual contents: face (facial expressions, gaze), body (limb, hand gesture, upper body), in the form of 
%-	Generating sound: voice prosody, rhythm, voice emotion, expressive music, improvisation, computer music … (some are beyond this paper)
%-	Supporting scene changes and camera cuts: mesh-up in concert video (beyond the scope of this paper)
%-	Purpose: 1) generating videos in scale 2) not to replace film maker, but to provide previsualization
%-	Movie
%-	Dance
%-	Music performance
%-	Instruction video

%\textbf{Body movement generation from audio}  
Several attempts have also been devoted to generate body movement from music. %The commonly seen topics of body movement generation from audio include generating body movements from music, generating gestures from speech, and generating dance from music. One approach for such generation problem is using the RNN-based model. 
\cite{shlizerman2018audio} used a recurrent neural network (RNN) to encode audio features and then a fully-connected (FC) layer to decode it into the body skeleton keypoints of both pianists and violinists. In \cite{kakitsuka2016choreographic}, choreographic movements are generated from music according to the user's preference and the musical structural context, such as the metrical and dynamic arrangement in music. %provides crucial information for the musical movement generation.
A recent work considered modeling the violinist's body movement with individual parts, including right-hand bowing, left-hand fingering, and the expression of the whole upper body \cite{liu2020body}. 
Another recent work on pianists' body skeleton generation \cite{li2018skeleton} also consider musical information including bar and beat positions in music. The model combining CNN and RNN was proven to be capable of learning the movement characteristics of each pianist. 

The method for assessing the quality of music-to-body-movement generation is still an open problem, mainly because the mapping from music to motion is not one-to-one. Subjective tests has been conducted, and the questions to participants are how reasonable and how natural the generated body movement behaves with music \cite{liu2020body}. A commonly-used objective metric is the distance between the predicted keypoints and the keypoints estimated by pose estimation algorithms in the original videos, the latter is taken as the pseudo-ground-truth \cite{shlizerman2018audio}. Recently, the accuracy of bowing attack is proposed specifically for evaluating the quality of virtual violinists' body movement. Another issue in this task is that there is no unique ground truth. In this paper, we argue that creating a testing data with multiple versions of ground truth (e.g., multiple performance videos for the same music piece) could represent a further step to analyze the result of the body movement generation models.

\section{System Overview}\label{sec:overview}
\begin{figure*}[t]
    \centering
    \includegraphics[width=0.85\textwidth]{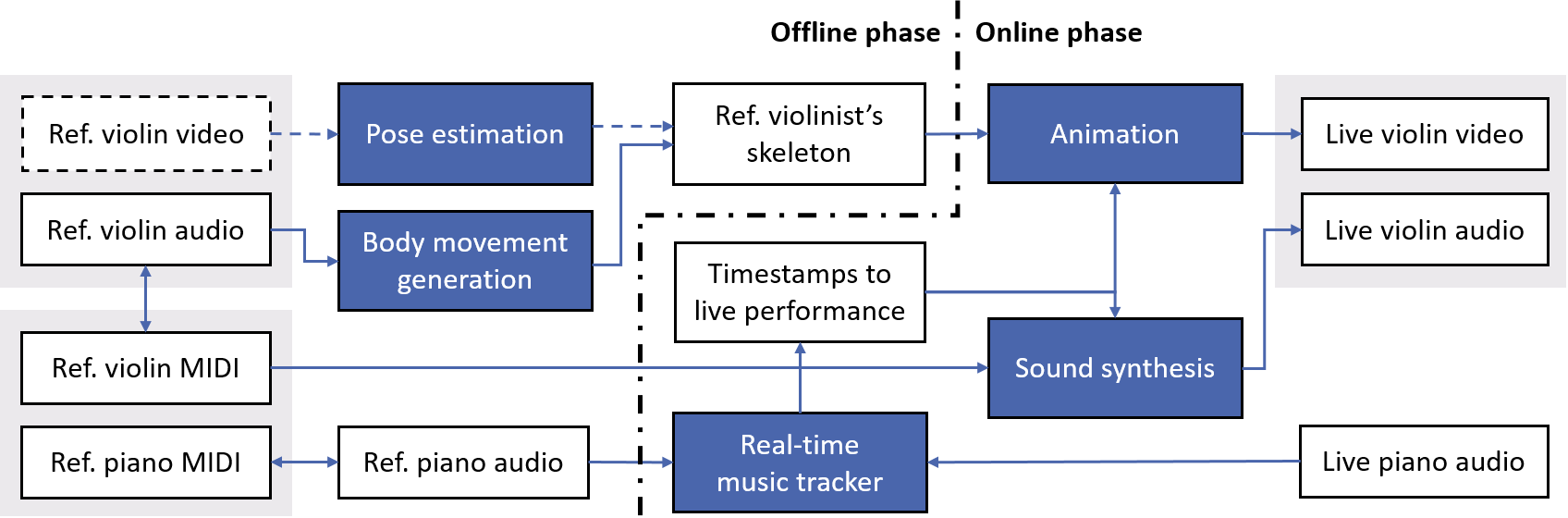}
    \caption{The proposed virtual musician system for animation and interaction. }
    \label{fig:system}
\end{figure*}
%There are many possible applications along with music performance. 

Figure \ref{fig:system} illustrates the system diagram of the proposed human-computer duet system. In the description below, we refer to the person who prepares the music pieces to be performed as a \emph{director}, and the person playing music with the virtual musician as a \emph{performer} or a \emph{human musician}. While it
is possible for a user to serve both of these roles, we distinguish them to clarify different working stages in our system. A MIDI or a MIDI-synthesized audio specifying exact note events is named as a \emph{reference}. A recording of a music performance made in the preparation stage (i.e., the offline stage) is referred to as a \emph{recorded} or \emph{rehearsed} performance. The performance happening in the scene is referred to as \emph{live} performance. % and the recording of a live performance is simply called the \emph{performed} recording. 

The proposed system is built upon the mechanism to synchronize altogether the reference, rehearsed, and live contents.
First, a director select a music piece to be performed. In our discussion, the music piece contains a reference violin track and a reference piano track, both of which are perfectly aligned with each other. To achieve better music tracking during performance, the performer may prepare a piano recording of her or his own. This recording pertains the performer's traits of music better than the reference MIDI does. %, and can help in synchronizing the live performance.
The reference and performed piano are aligned using the offline DTW algorithm such that the temporal correspondence between both recordings can be retrieved directly. In live performance, the real-time music tracker follows the live piano played by the performer, and by the online DTW (ODTW) alignment, it monitors the current position on the reference tracks. The music tracker then triggers the visual animation and sound synthesis items to generate the body movement and sound of the virtual musician's live violin performance at the current position. 

As for automatic animation, there are two possible ways to generate the virtual violinist's motion. The basic way is employing pose estimation \cite{pavllo20193d} to extract pose sequence on a violin video recording of the selected music piece for animation. The advanced way is applying music-to-motion generation techniques, where the motion generation model is trained on a music video dataset of violin performance. With this model, a pose sequence can be generated from the reference violin audio, which is aligned with the reference violin MIDI in order to directly synchronize with the piano signal in live performance. The generated body skeleton is bound with a 3D human rigged model and a violin model for character animation. 
%The reference violin and piano audios are both optional in the proposed system; the system still works when such recordings are unavailable, as the live recordings can be directly aligned with the reference ones, and violinist's motion can also be generated directly from the reference recording. 
According to the real-time tracking result with the live piano audio, the virtual musician's audio and video parts corresponding to the performance are rendered with a real-time sound synthesizer and an animator. The details of these building blocks will be introduced in the following sections.

\section{Real-time Music Tracker}
\label{sec:music_tracker}

\begin{figure}[t]
    \centering
    \includegraphics[width=0.45\textwidth]{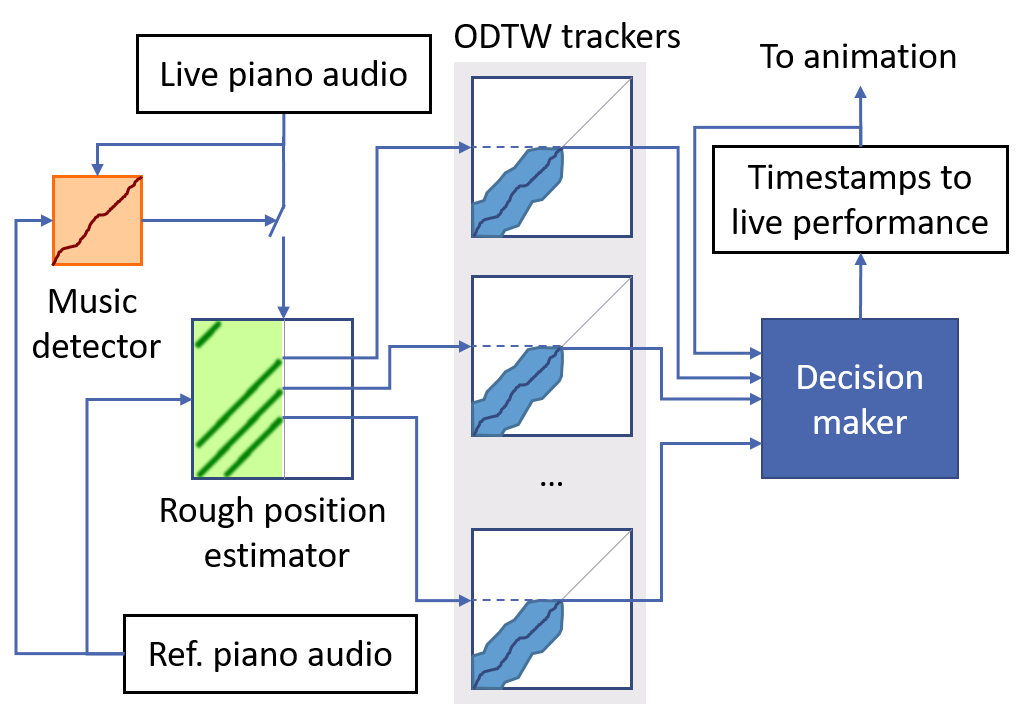}
    \caption{The real-time music tracking system}
    \label{fig:track}
\end{figure}

%The real-time music tracker is based on ODTW to synchronize the live piano and the rehearsed violin. Besides ODTW, conventional DTW is also employed in this paper to synchronize the reference and rehearsed data in the offline stage. 
In this section, we review the DTW and ODTW algorithms, and describe the implementation of the real-time music tracker. %\emph{performance audio} $x^{\text{pn}}_i$ and the \emph{reference audio} $x^{\text{pn}}_m$. 
%The reference audio can be the audio directly synthesized from MIDI, or a previous recording by the human performer $x^{\text{pn}}_h$. Our pilot study shows that tracking the recording can achieve better tracking accuracy.

%which is the main alignment algorithm in the music tracker. Then details of the music
%tracker are addressed.

\subsection{Dynamic time warping}

Given $\vX:=\{x_p\}^P_{p=1}$ and $\vY:=\{y_q\}^Q_{q=1}$, the features extracted from any two sequences mentioned in Section \ref{sec:overview} and Figure \ref{fig:system}.  %\textcolor{red}{[mention audio processing details in some other place]} %where in this work $\vX$ is the chroma sequence of the reference performance, and $\vY$ is the chroma sequence of the live performance. 
%$\text{\text{\tmtextbf{\tmtextit{\mathrm{}}}\tmtextbf{X}}=\tmtextbf{\text{x}}\tmrsub{1},
%\text{\tmtextbf{x}}\tmrsub{2}, {\ldots}, \tmtextbf{x}\tmrsub{\tmtextit{m}}}$
%and
%$\text{\text{\tmtextbf{\tmtextit{\mathrm{}}}\tmtextbf{}\tmtextbf{Y}}=\tmtextbf{\text{y}}\tmrsub{1},
%\text{\tmtextbf{y}}\tmrsub{2}, {\ldots}, \tmtextbf{y}\tmrsub{\tmtextit{n}}}$,
%where each \tmtextbf{\text{x}}\tmrsub{\tmtextit{i}} and
%\tmtextbf{\text{y}}\tmrsub{\tmtextit{j}} 
Each $x_p$ and $y_q$ are the $p$th and $q$th feature vectors of $\vX$ and $\vY$, respectively. %and $M$ and $N$ are the length of the reference performance and the live performance respectively, 
%$1\leq p \leq P$ and $1\leq q \leq Q$.
The DTW algorithm finds the path $\mathcal{W}:=\{(p_k, q_k)\}^{K}_{k=1}$, $1\leq p_k \leq P$ and $1\leq q_k \leq Q$, such that the total alignment cost $\sum^K_{k=1}d\left(x_{p_k},y_{q_k}\right)$ is minimized, where $d\left(x_{p_k},y_{q_k}\right)$ is the Euclidean distance between $x_{p_k}$ and $y_{q_k}$. % to represent local alignment cost.  
%$\tmmathbf{\text{X\text{}}}$ and $\text{\tmtextbf{Y}}$respectively. 
%The DTW algorithm takes two sequences as input.
%The output of DTW is an alignment path\tmtextit{
%$\text{\tmtextbf{W}} = W_1, W_2, \ldots, W_l$}, %where $l$ is length of the
%path. Each $W_k$ in $\tmmathbf{W}$ is a pair $(i_k, j_k)$ that means
%$\text{\tmtextbf{x}}_{i_k}$ and $\text{\tmtextbf{y}}_{j_k}$ are matched at
%$k$-th position in the alighment path. The target of DTW is to find the path
%$\tmmathbf{W}$ such that tocal alignment cost $\underset{k =
%1}{\overset{l}{\sum}} d \left( \mathit{\text{}} \text{\tmtextbf{x}}_{i_k},
%\text{\tmtextbf{y}}_{j_k} \right)$ is minimum, where $d$ is a distance measure
%between each pair of $\left( \text{\tmtextbf{x}}_i, \text{\tmtextbf{y}}_j
%\right)$ as the local cost function. 
The path $\mathcal{W}$ therefore represents the optimal mapping from $\vX$ to $\vY$. 
%Conventional DTW imposes two constraints on the path $\mathcal{W}$. First, $\mathcal{W}$ starts from $(1, 1)$ and ends at $(P, Q)$. Second, the alignment path goes monotonically with a fixed set of unit steps. Generally, for each $(p_k, q_k)\in\mathcal{W}$, $(p_{k+1},q_{k+1})$ is chosen only from $(p_k + 1, q_k)$, $(p_k, q_k + 1)$, or $(p_k + 1, q_k + 1)$. In other words, we can only take one of $(0, 1)$, $(1, 0)$ or $(1, 1)$ as the step from $W_k$ to $W_{k + 1}$. Under these constraints, 
The optimal alignment path can be found by dynamic programming. 
%First, we calculate an $m \times n$ cost matrix $\vC$, where $\tmmathbf{C} (p, q)$ stands for the distance measure $d \left(x_p,y_q\right)$. 
The cumulative cost matrix $\vD\in\mathbb{R}^{M\times N}$ is calculated recursively as follows:
\begin{equation}
  \vD\left[p_k, q_k\right] = d \left(x_{p_k},y_{q_k}\right)+\min
  \begin{cases}
    \vD \left[p_k - 1, q_k\right]\\
    \vD \left[p_k, q_k - 1\right]\\
    \vD \left[p_k - 1, q_k - 1\right]\,.
  \end{cases}
\end{equation}

$\vD\left[p_k, q_k\right]$ stands for the minimum of total cost of the alignment path from $(1, 1)$ to $(p_k, q_k)$. After $\vD$ is calculated,
we backtrack the matrix from $\vD \left[P, Q\right]$, find each optimal step
iteratively and get the optimal alignment path $\mathcal{W}$.

%\subsection{Online dynamic time warping (ODTW)}

%ODTW is a variant of DTW. 
The conventional DTW assumes that the whole recordings of both sequences are known in advance. 
%In traditional DTW, the alignment path is obtained by back tracking over the cost matrix. (i.e., $\vX$ and $\vY$)
This is however not the case in live performance, for it is impossible to retrieve the performance content in the future. 
%has some disadvantages in realtime applications, in which on-line sequence alignment is necessary. DTW requires that the inputs are known in advance, but the sequences are extendable in realtime applications. 
Besides, the quadratic complexity of DTW limits the algorithm from real-time computation. The ODTW algorithm \cite{dixon2005live} is proposed to solve these problems with two main modifications: first, instead of computing the cost matrix $\mathbf{D}$ from all $(p,q)\in\left[1,P\right]\times\left[1,Q\right]$, the time intervals considered in computing $\mathbf{D}$ is dynamically determined by the warping path direction and a \emph{search depth} with a fixed length $c$. Second, instead of finding the optimal path $\mathcal{W}$ in the backtracking process after knowing the complete cost matrix, the path at the $(k+1)$th time step is determined right after $\mathbf{D}\left[p_k,q_k\right]$ is known. %In other words, the process of cost matrix accumulation and the process of path assignment are performed alternatively.
This is controlled by another parameter, which resets the direction of warping path when it has been stuck into one direction for a certain time steps.  
%is defined in the algorithm, which reduces the computation complexity to constant in each iteration. The other is that the cost matrices are computed incrementally. Then the alignment path is extends forward accordingly. These modifications make ODTW applicable in realtime usages.
%The followings are brief introduction of ODTW process. The cost matrices $\tmmathbf{C}$ and $\tmmathbf{D}$ are defined as the same in the conventional DTW. In each iteration, the direction to expand the cost matrices is decided by the values in the front row and column of $\vD$. After new parts of the matrices are calculated, the alignment path is extended in the same way. As the input sequences of ODTW growing, these procedures are repeated to update the alignment to the newest input.  There are also some extensions that can improve accuracy of the algorithm, such as backward-{forward} strategy[?] and tempo models[?].
%at time $(p_k,q_k)$, the ODTW algorithm deals with the following three cases: 1) if the minimal cost is at $(p^\prime_k,q_k)$ where $p^\prime_k\leq p_k$, then compute $\vD_{p_k-c+1:p_k,q_k+1}$ by means of Equation 1 and assign the $(k+1)$th step of the warping to the index which has minimal cost at the $q_k+1$;2) A fixed search depth $c$ is introduced. Every step the accumulative cost matrix $\vD(p,q)$ is calculated at at new step $p+1$, the cost is accumulated from $(p+1,q-c+1)$ to $(p+1, q)$ by applying a parameter $c$.
As a result, ODTW achieves the performance in linear time and space. See \cite{dixon2005live} for the details of the ODTW algorithm.

\subsection{Implementation details}

The real-time music tracker we employed is a re-implementation of the \emph{'Any time' music tracker} by Arzt and Widmer \cite{arzt2010simple,arzt2010towards}. The tracker operates in multiple threads, where each thread is in charge of a sub-task. The sub-tasks include a music detector, a rough position estimator, several ODTW trackers and a decision maker, as shown in Figure \ref{fig:track}. These components enables an automatic way to trigger the tracking mechanism when the performance begins, and track the performance continuously with the computing cost affordable for a laptop. %In the following, we first introduce the scheme of the music tracker. Then the components are addressed respectively.
The audio features of both the live and the rehearsed piano signals are the rectified spectral difference, which is the first-order difference of spectrum with the negative elements set to zeros. The feature is derived from the log-frequency spectrum which frequency resolution is one semitone and time resolution is 20ms. %Such a spectrum is derived from the STFT with a Hamming window of size 46ms and with a hop size of 20ms. 
%By comparing every two consecutive frames, only the temporally increasing spectral components are preserved and other components are set to zero \cite{arzt2010towards}. 
Two types of features with low and high temporal resolutions are employed. %for the rough position estimator and the ODTW trackers, respectively: 
The high-resolution feature for the ODTW trackers is the above-mentioned spectral difference, and 
%The audio are transferred to feature bins with semitone interval using spectrum derived by Fourier Transform. 
%For each frame we calculate the difference with the previous frame and only reserve positive differences to emphasize note onsets, as the suggestion in \textcolor{red}{[?]}. 
The low-resolution feature for the rough position estimator is a downsampled version of the high-resolution spectrum. % is the high-resolution feature smoothed with a Hann window with size 600ms, %STFT with window size of 600ms and hop size of 300ms. The low-resolution feature is derived from the high resolution feature by convolution with a Hann window with length 30, 
%and down-sampled by a factor of 15. Both the features are normalized to sum up to one. \textcolor{red}{Struture of the system is shown in Figure 1.} %\textcolor{red}{(what kind of normalization?)(how to compute the low-resolution feature in real time?)}.

The music detector identifies the time instance that the performance starts and tells the
system to start tracking. The music detector is implemented in a simple yet practical way: we use the conventional DTW to align the live audio with the first 0.5 sec of the rehearsed audio. If the alignment cost is lower than a preset threshold, the music detector triggers the music tracking process. %considers that the play has started.
%Though the approach is simple, it is practical in real world performance.

%However, for concert applications, we also need to know where the play of music is on the score in realtime. To achieve this, we use an audio file synthesized from a MIDI file as reference. The MIDI file is a representation of the score. Therefore, we can get score position according to the MIDI file during real-time audio alignment.

%In the live performance, new audio frames come in the music tracker continually. 
%The music detector needs to decide whether and when the performance starts\textcolor{red}{, and if the performance starts, it triggers the real-time music tracker at the position where the performance starts.}

The rough position estimator returns a set of possible current positions the live piano signal corresponding to the rehearsed piano. This is done by computing the similarity (measured by the Euclidean distance) of the latest nine seconds of the live audio to the segments of the rehearsed audio ending at any positions. %, with a solution of 600ms. 
The positions which reach high similarity are taken into the set of possible current positions.  
%are aligned with the whole reference audio using the low resolution features, using a \textcolor{blue}{local backward greedy alignment algorithm} \textcolor{red}{(what is it?)}. From the alignment results, 5 percents of the positions with lower costs are selected and passed to the decision maker. 
In \cite{arzt2010towards,arzt2010simple}, the rough position estimator keeps all possible positions to deal with cases where the music might jump to any position on the score in cases like practice or improvisation. 
In our case, since we use a \emph{rehearsed} audio, such jumps are unlikely to occur. We keep using the rough position estimator just for faster correction when the ODTW trackers have wrong alignment.

In the tracking process,  the ODTW algorithm is utilized to align the high-resolution features of live piano and the rehearsed piano. Multiple ODTW threads works in parallel, and each thread deals with a possible current position estimated by the rough position estimator.
%The \textcolor{blue}{ODTW trackers} run the main matching algorithm in the music tracker. 
%These matchers align the real-time audio frames to the reference audio by ODTW in parallel, starting from the possible positions assigned by the decision maker. 
In addition to the precise estimation of current position, the cumulative alignment cost and instantaneous tempo of the live audio is also recorded. %These matchers are \textcolor{red}{also} used for calculating tempo of music. %Forprecise alignment results, the matchers use the high-resolution features of 20ms. 
%The alignment results, including time positions, tempos and costs 
These results of all the ODTW threads are then fed into the decision maker for final decision of the current position. In practice, we use about 2-4 ODTW threads in the system, depending on the available hardware resource.

Finally, the decision maker selects an ODTW thread as the trusted one and output the tracking result of that selected ODTW thread. This is done by finding the minimal cumulative cost value of the currently selected ODTW threads to all the other ODTW threads. 
%manages these positions and verify them by the parallel ODTW matchers. Then the decision maker choose the most
%credible ODTW matcher and take the alignment result as output. 
%The decision maker manages the music tracker. 
%At each new time step, any new possible position given by the rough position estimator is assigned to an idled ODTW thread to trigger a tracking process at that position. Then, the decision maker collects the tracking results from all running ODTW matcher and decides which matcher is credible. 
%While the alignment results from the matchers updates, the decision maker verifies these matchers by calculating cumulative cost differences between the verified matcher and the selected credible matcher. 
%Other untrusted ODTW threads are deactivated and are triggered again when new possible positions are found. 
As new audio frames received, the decision maker repeats these procedures and updating output with the credible ODTW matcher.
%\textcolor{red}{Thresholds of cumulative difference are set to assign a newly trusted matcher and deactivate those matchers with bad alignment results. Those idled matchers are triggered again when new possible positions are found. As new audio frames feeds in, the decision maker repeats these procedures and updating output with the credible ODTW matcher.}
The real-time music tracker is implemented in Python 3.7 with the \texttt{multiprocessing} package for parallelization.

\vspace{-0.2cm}

%\subsection{Auto Accompaniment}

%The accompaniment is defined as the counterpart of the performance. A piece of music may consists of several \tmtextit{voices}, including the plays by different instruments like piano, violin, etc. 

%In our AI concert system, the performance may contains only part of these voices, and the rest is the accompaniment. Contents in the performance and the accompaniment is assigned in advance. 

%The target of auto accompaniment is to detect the performance and play the accompaniment accordingly. More specifically, we have to know the position of the performance on the score, get the corresponding accompaniment, and play the accompaniment synchronizing to the performance. In this section, technical details of auto accompaniment are addressed.

%The accompaniment data is represented by a MIDI file. The MIDI file contains
%score information and matches the performance. Notes in the accompaniment are
%recorded as events in the MIDI file. During the performance, the position of the performance is known with the music tracker, and therefore we can play the accompaniment notes accordingly. 

%We use a synthesizer (Ableton Live) to play the accompaniment in realtime. To play the accompaniment notes, we just need to find the correponding events in the MIDI file and send the messages to the synthesizer to generate accompaniment audio.

\section{Body movement generation}
\label{sec:body}

As discussed in Section \ref{sec:overview}, the virtual musician's body movement for music performance is obtained with two approaches: pose estimation and %from existing videos in which human violinists performed the music, or automatic 
body movement generation. %from the music to be performed. 
The latter approach requires a dataset with audio and pose contents of violin performance to model the music-to-body-movement correspondence. 
For pose estimation in both the approaches, we adopt \cite{pavllo20193d} to extract the 3-D position of the violinists’ 15 body joints, resulting in a 45-D body joint vector for each time frame. The joints are extracted frame-wisely at the video’s frame rate of 30 fps. All the joint data are normalized such that the mean of all joints over all time instances is zero. The normalized joint data are then smoothed over each joint using a median filter with a window size of five frames.

\begin{figure}[t]
\centering\includegraphics[width=0.95\columnwidth, trim={0cm, 1.2cm, 0cm, 0cm}, clip]{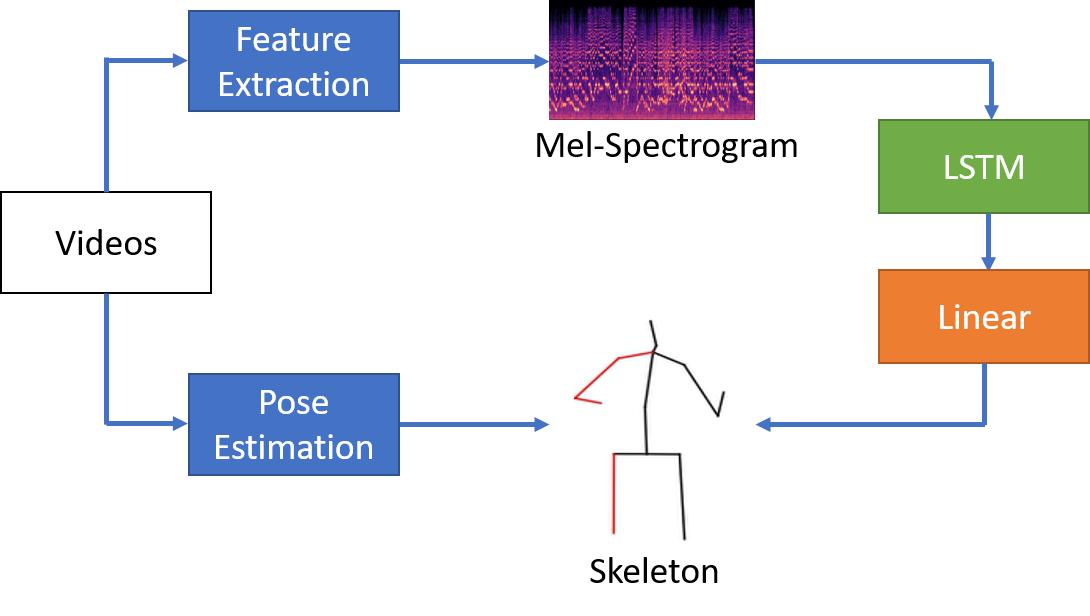}
\caption{The overview of body movement generation.}\label{body_movement}
\end{figure}

For body movement generation, we extend the framework of audio-to-body (A2B) dynamics \cite{shlizerman2018audio}, a 2-D body movement generation framework, into a 3-D body movement generation framework, as illustrated in Figure~\ref{body_movement}. 
%, we introduce a generation model which is utilized to predict playing movement on violin or piano music proposed in audio-to-body dynamics [],
%but we here apply it on 3-D skeletal prediction to be an extended application. 
The framework is constructed by a RNN with long-short-term memory (LSTM) units plus a FC layer. The input audio features extracted from the recorded violin are first fed into the RNN. The output values of the RNN are then fed into the FC layer. To model the long-term dependency between music and body movement, a delay mechanism between the input and output is introduced \cite{shlizerman2018audio}: the model takes input audio feature at time $t$ and outputs the skeleton data at $t-\eta$. In this paper, we set $\eta=6$ frames. %of is delayed from the input sequence of audio features with six frames .
%with time delay and a fully connected layer with dropout. We follow their  setup and use MFCC feature as input.

%\subsection{Data pre-processing}

%\textbf{Audio feature extraction}
The input audio features are the 128-D mel-spectrogram with a temporal resolution of 1/30 sec. In this way, the audio and skeletal features for training are synchronized. All the features are extracted with the \texttt{librosa} library~\cite{mcfee2015librosa}.
%We use the librosa library [] to extract audio features.  Each music track is sampled at 44.1 kHz.  We conduct STFT with a sliding window with length of 4096 samples and a hop size of 1/30 sec, which fits the to frame rate of the video sequence. The resulting feature vector  
%\textcolor{red}{As the violinist's body movements correlates with musical beats, all the training data are segmented on the barlines according to the beat positions and time signature.} 
Each training sample is segmented at the downbeat positions, as we consider such structural information as a guide of the corresponding body movement. 
%the violinist's body movement correlates with musical beats, the segmented data should include musical structure. 
To do this, we first annotate beat position on the MIDI file of each musical piece, and then use DTW to align beat position between the MIDI-synthesized audio and the recorded audio performed by human violinists.
%by shifting one barline with fixed length. In this paper, 
As a result, each training sample starts from a downbeat and is with length of 300. All the segmented data are normalized by z-score. %, and 20 \% of the training data would be split as validation data further. 
%Since the tempi and time signatures of the musical pieces are different, 
%we zero-pad every segmented sequence to the maximal length of segmented sequences in order to keep the lengths of all training samples being the same.
The model is trained by minimizing the $L_1$ loss between the output and the ground-truth skeleton sequence. The hidden unit of LSTM is 200, followed by dropout 0.1, and the model is optimized by Adam with $\beta_{1}= 0.9$, $\beta_{2}= 0.98$, $\epsilon=10^{-9}$, and learning rate is set to 0.001. %Each fold would train 100 epoch and we choose inference model which has the lowest validation loss in the training procedure.

\section{Model binding and animation}\label{sec:integration}

The next step to launch the animation is to bind the generated skeleton data to the 3-D human rigged model (called \emph{skeleton binding}) and also to the violin model (called \emph{violin binding}). %The two tasks are described as follows. 
The model binding and the animation processes were done with the cross-platform game engine, Unity 3D 2018.3. %\footnote{https://unity.com/}  %, while the motion data stream is in the format of Kinect V2 \cite{wiedemann2015performance}. %As the motion dataset and the animation are from different system, such inconsistency could be usually met. Therefore
%Since the joint coordinate of the two types of skeletons are defined in different ways, we use the the built-in utility ``Global Position'' in Unity 3D %a method 
%to fit the skeleton data %(either generated by the body movement model or captured from video data, as mentioned in previous sections)
%to the 3D human rigged model. %is required. 

% In addition to the inconsistency of the number of joints and joint positions between Humanoid and Kinect V2, the way the two models to define the coordinates are also different. The joint coordinates in Humanoid are hierarchical: every joint has its own coordinates, and the child joint takes the parent joint as the origin of coordinates. All the joints in Kinect V2 share the same coordinate system, and it origin is the Kinect V2 Sensor itself. Such inconstancy could be solved with the built-in utility “Global Position” in Unity 3D. 

\subsection{Skeleton and violin binding}\label{subsec:binding}
We use the Humanoid\footnote{https://docs.unity3d.com/Manual/AvatarCreationandSetup.html} provided by Unity as our 3-D human rigged model. The skeleton data obtained from either pose estimation or body movement generation are in very limited resolution, and with noisy movement and sporadic articulation. To avoid unstable body movement during animation, we consider applying the inverse kinematics (IK) method to bind the motion data with the rigged model with refined skeleton joint positions.
%2) the motion data are noisy because of the limitation of accuracy in pose estimation technology. Therefore, IK is preferable in this stage.
%There are two possible ways to bind the motion data with the rigged model: \textcolor{red}{forward kinematics (FK)} and inverse kinematics (IK). There are trade-offs between the two methods. In comparison to IK, FK displays the true motion data of every joint on the 3D rigged model, but it is inevitably more sensitive to the instability in the data. On the other hand, 
In comparison to forward kinematics (FK) which directly assigns all the skeleton joint rotation values to the model, IK allows one to infer the whole skeleton joint data from only the skeleton joints which are either more significant for violin performance (e.g., the upper body and hands), or better estimated by the pose estimation or body movement generation models. 
%take only a few joints and infer other joint data through the 
The resulting skeleton data obtained from the IK equations therefore become more smoothed and reasonable. %therefore resulting in smoothed and reasonable skeleton joint positions. %but might introduce errors. 
%Considering that our motion data tends to be noisy (as it is from pose estimation and the body movement generation model), and also considering that violinist’s body movement is basically based on only a few joints at the upper body and the hands, we adopt the IK method in order to get more stable animation. 
In this system, we consider four joints for IK inference: left shoulder, left hand, right shoulder, and right hand. The base of the spine is taken as the center of body rotation. The system is implemented with the Full Body Biped IK algorithm\footnote{http://www.root-motion.com/finalikdox/html/page8.html} in \texttt{Final IK} 1.9\footnote{https://assetstore.unity.com/packages/tools/animation/final-ik-14290}, which is provided by the Unity Assets Store.

\begin{figure}[t]
    \centering
%    \subfigure[]{
    \includegraphics[width=\columnwidth,]{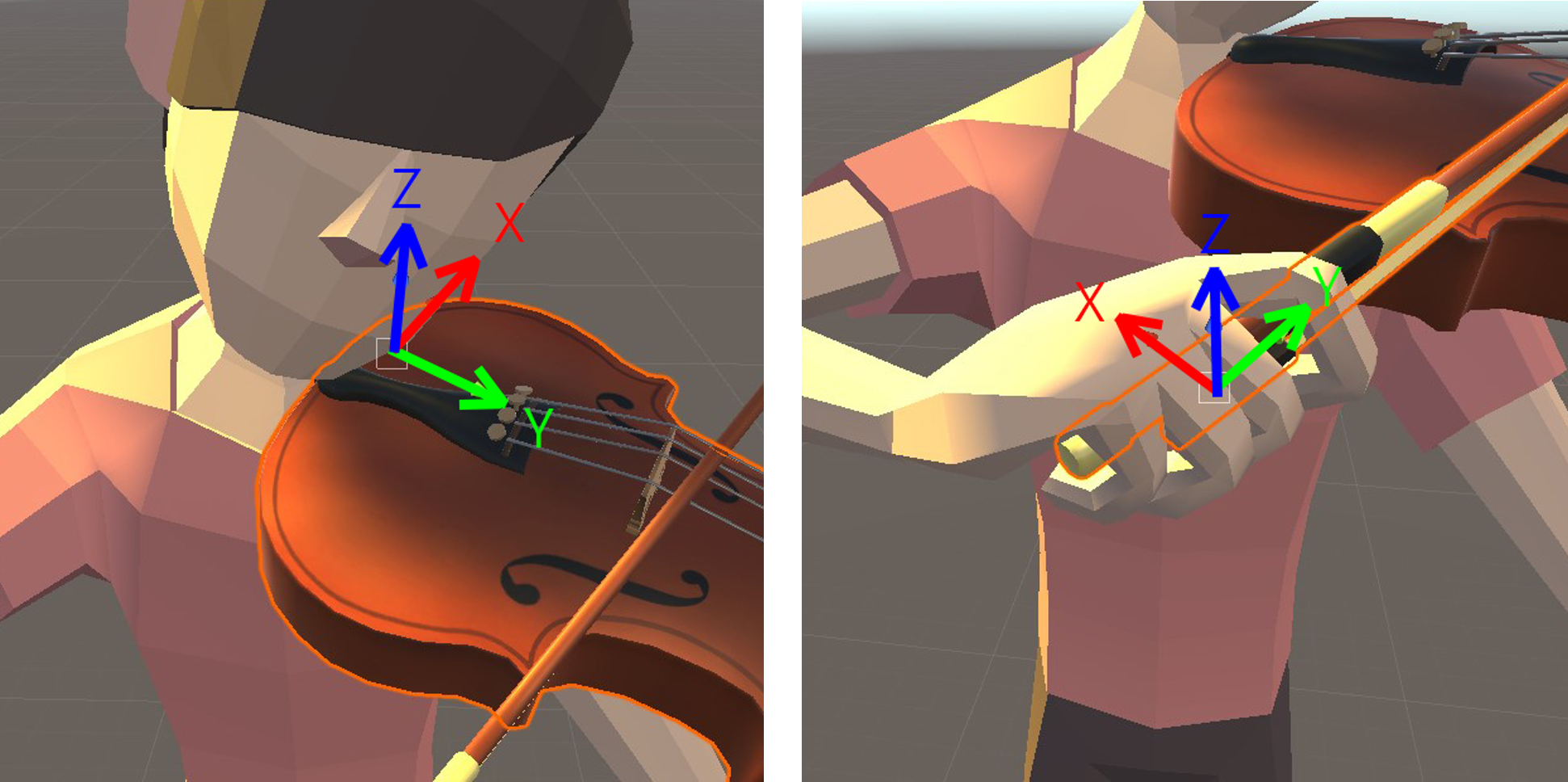}
%    }
%    \subfigure[]{
%    \includegraphics[width=0.8\columnwidth]{fig%/figure3-b.jpg}
%    }
    \caption{The coordinate setting for violin binding. Left: the violin body. Right: the bow.}
    \label{fig:violin_binding}
\end{figure}

%\subsection{Violin binding}
The violin model has two parts: the violin body and the bow. Figure \ref{fig:violin_binding} illustrates the coordination of the violin model. For the violin body, the origin of coordinates is at the bottom of the chinrest, with the $y$-axis toward the scroll, and the $z$-axis perpendicular to the frontal plane of the violin body. The origin of coordinates is placed in the middle of the violinist’s left shoulder and neck, making the $y$-axis of the violin body be connected to the left hand joint. For the bow, the origin of coordinates is at the bow frog and is connected to the right hand joint. The bow stick is placed in the $y$-axis. The $y$-axis of the bow is also connected to a position in between the bridge and fingerboard of the violin body. The position and angle of the violin body and bow are updated every time after the skeleton is updated with IK.

\subsection{System integration}

\begin{figure}
\centering
\includegraphics[width=\columnwidth,trim={0cm, 0cm, 0cm, 1cm}, clip]{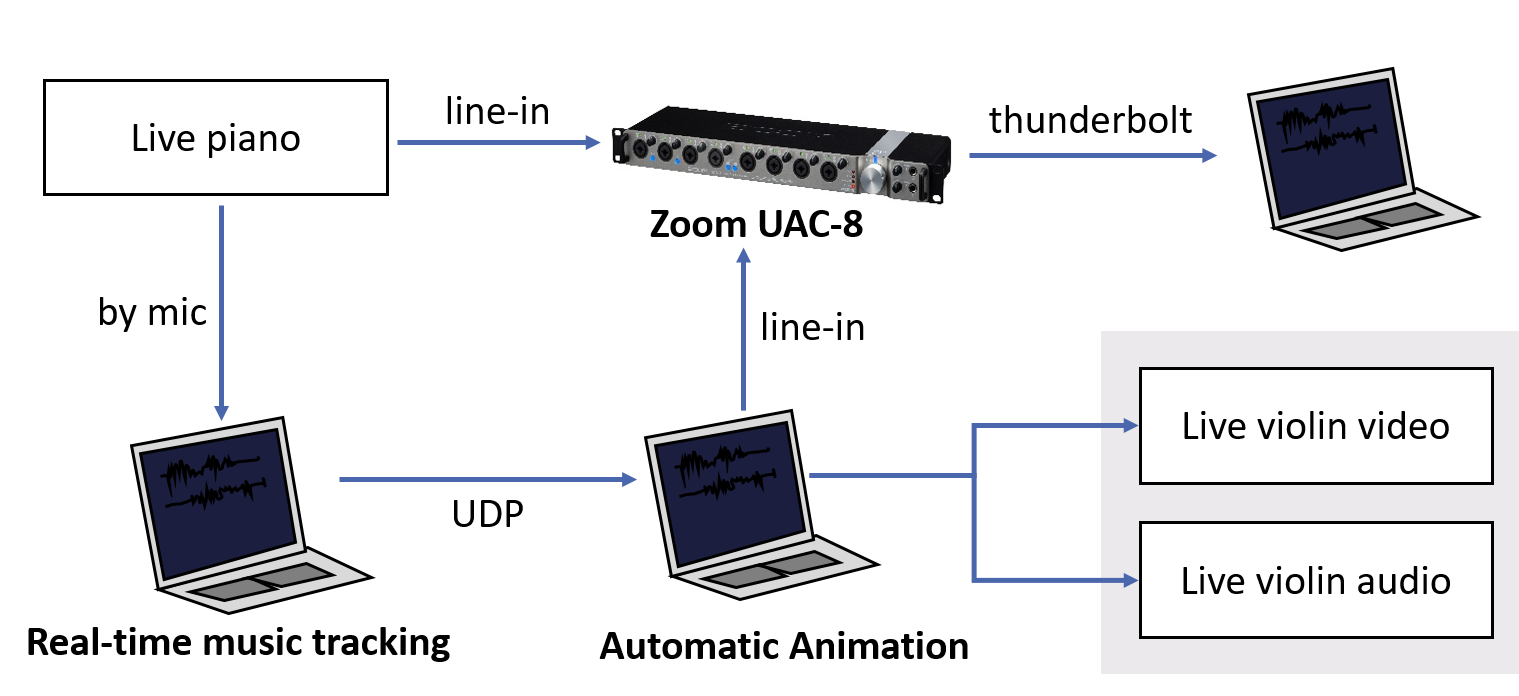}
%\hspace{0.3cm}
\caption{The distributed experiment setting of the computer-human duet system. The two laptops shown below are for performance, and the third laptop shown above is for recording only.}
\label{fig:recording}
\end{figure}

%All of the above-mentioned building blocks are implemented independently, and can be all executed on a single machine, or be distributed into different machines. 
To ensure system stability for real-world performance, we distribute the on-line computing counterparts described in Section \ref{sec:music_tracker}, \ref{sec:body}, and \ref{subsec:binding} into two laptops: one laptop takes charge in real-time music tracking, and the other one takes charge in automatic animation; see Figure \ref{fig:recording}. The two laptops communicate with each other through a local network: during a performance, the real-time music tracker continually sends the newest tracking result (i.e., the corresponding time of the rehearsed piano audio that matches the live piano), which is packed in JSON format, to the automatic animator on the other laptop through the User Datagram Protocol (UDP). The skeleton data generated from the rehearsed violin and the MIDI of reference violin are both pre-stored in the automatic animator. We use Ableton Live to play the sound, and Unity to display the visual content of violin performance. 
The real-time music tracker sends a UDP package to the automatic animator every 20ms. The automatic animator receives it and manages two tasks. First, it guides the sound synthesizer to play the violin performance from MIDI, according to the received tracking results, such as to make the violin performance synchronized with the live piano. Second, it guides the animator to display the virtual violinist’s body movement, which also follows the newest tracking result. %in real time. The goal is to make the resulting animation in harmony with the live performance. 

%it is to synchronize the virtual musician's behaviors with the human-performed music. In practice, the latency of the local network is short enough and negligible in comparison to other components such as the music tracker that introduces larger latency. 

%The real-time music tracking system gives the timestamp of the reference score corresponding to the timestamp of the live performance. The $M_t$ is then sent to the virtual musician animator through the UDP protocol. 

%The virtual musician animator receives the tracked $M_t$ every \textcolor{red}{XX ms}. %Then, the motion data at $M_t$, $M_data$, is taken for animation. 
\section{Results}
\label{sec:result}
\begin{figure}[t]
\centering
\includegraphics[width=\columnwidth, trim={0cm, 0cm, 0cm, 1cm}, clip]{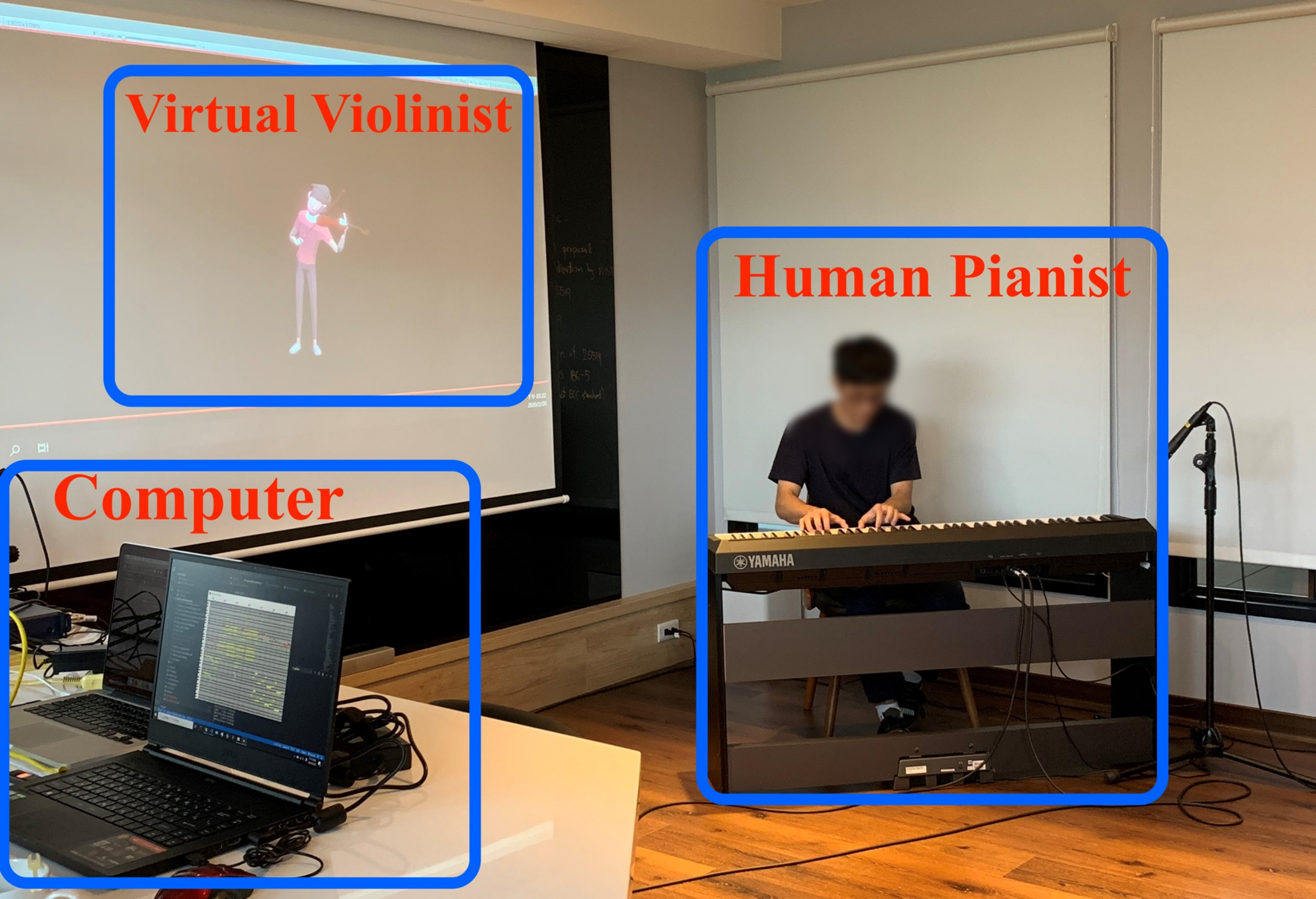}
\caption{The setting of the human-computer duet system. Note that to verify the robustness of the real-time music tracking unit, the sound of live piano for music tracking is collected by a microphone (at the right of the photo) rather than collected by line-in.}
\label{fig: Concert}
\end{figure}

%\subsection{Demos and concerts}

\begin{figure}[t]
\centering
\includegraphics[width=0.97\columnwidth]{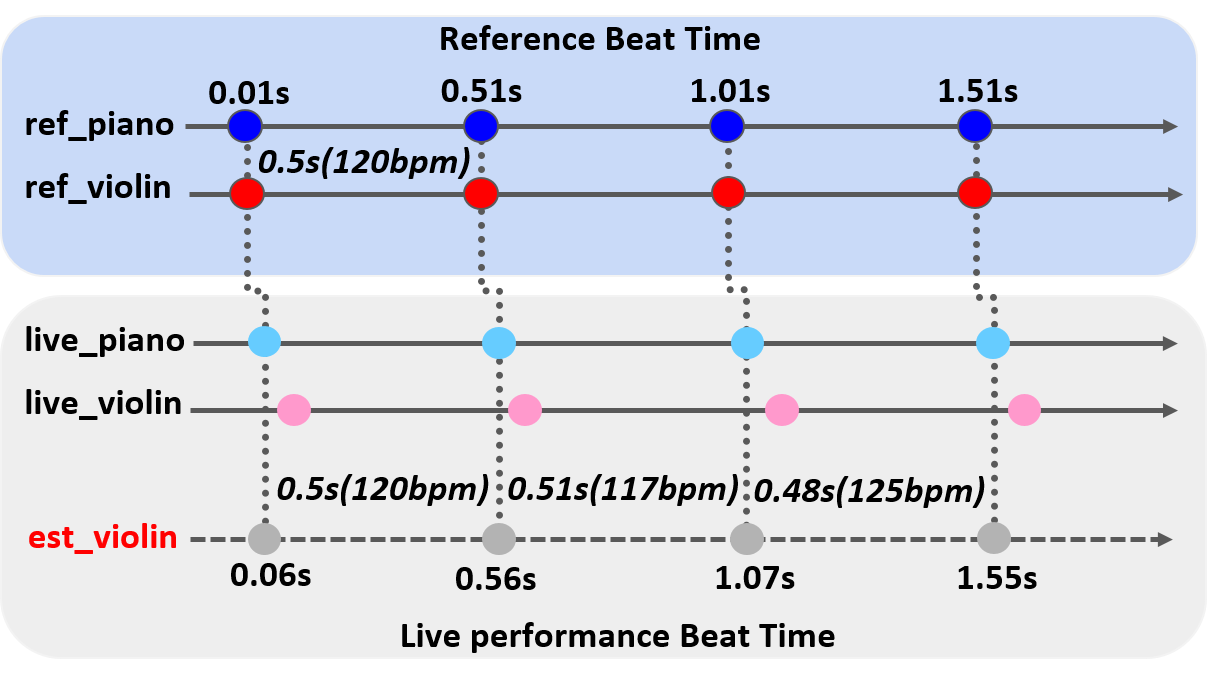}
\caption{A conceptual illustration of evaluating the real-time music tracking system on a sample in 120 BPM. Circles denote beat positions, and the circles connected with a vertical dashed line are synchronized. Note that the synchronization result determines how the violin part should be played (i.e., the estimated violin part).}
\label{fig: offline_DTW}
\end{figure}

The proposed human-computer duet system can be properly set in a performance venue, as shown in Figure \ref{fig: Concert}. The system has been publicly demonstrated twice. The first one was at the faculty lounge of the authors' Institute, where around 20 people watched the demo. The second one was in a public, ticket-selling concert held in a famous experimental music venue located in the authors' city, where 120 people attended the concert. Detailed information of these activities are hidden for double-blind review. The music piece we chose to perform and test on is \emph{Violin Sonata No.5, Op.24 (Spring), movement I. Allegro} by Ludwig van Beethoven. The system works stably over the whole performance, which lasts for around 10 minutes.
%The time signature of the \emph{Spring} sonata is 4/4 %(four beats per bar, with each beat being a quarter note) 
%and the tempo of the reference score was originally set to 120 bpm. 
% In other words, the duration of one beat, a quarter note, is ideally 0.5 seconds (500ms). 
The recordings of the public demo and concerts are attached in the supplementary materials. 

We received the audience's responses saying that the performance was insightful and they would like to join in again if similar kinds of concerts will be held in the future. However, we still regard the objective evaluation as a more precise way to assess the quality of the system. In the followings, we describe our proposed quality assessment approach, and discuss the ways to improve the system based on the assessment results. 
%to measure the accuracy of a real-time music tracking system.

\subsection{Assessment of real-time music tracking}

We propose a novel method to evaluate the performance of real-time music tracking from real-world performance. The idea is to evaluate the deviation between online music tracking and \emph{offline alignment with the live recording}.
%In this section, we explicate the concrete steps and details of how we conduct the experiment to evaluate our music tracking system’s performance. We first show our recording setting (Fig. \ref{fig:recording}) and recapitulate briefly the real-time tracking, automatic accompaniment process and review terminology to further explain our systematic approach. Then we show our experiment result in terms of how we human perceive latency in music. 
The process is described as follows: %done with the following process:

\begin{enumerate}
    \item Record the live piano and live violin audio with a multi-channel recording device; see Fig. \ref{fig:recording}. Confirm that the reference piano (blue circles in Figure \ref{fig: offline_DTW}) MIDI and reference violin MIDI (red circles in Figure \ref{fig: offline_DTW}) are perfectly synchronized. 
    \item Obtain the time mapping between the recorded live piano and the reference piano MIDI with offline synchronization (see the dashed lines connecting the blue circles and the light blue circles in Figure \ref{fig: offline_DTW}). According to such a temporal mapping, one can synthesize the \emph{estimated violin} signal, which is the `expected' violin performance with the live piano; see the gray circles in Fig. \ref{fig: offline_DTW}. The estimated violin is taken as the benchmark for a music tracking system. 
    \item Obtain the time mapping between the live violin and the estimated violin with offline synchronization. If the time $t_i$ of the live violin is mapped to $t_j$ of the estimated violin, then the \emph{latency} of the real-time music tracking system performing the music piece at $t_i$ is defined as $\Delta[t_i]:=t_j-t_i$.  
\end{enumerate}

In other words, positive latency values mean that the music tracker drags and negative latency values means that the music tracker rushes. %The \emph{average latency} for an entire music clip with length $T$ is therefore $(1/T)\sum_t\Delta[t]$. 
Besides the instantaneous latency values, we also consider the \emph{average deviation} of a music piece, defined as $(1/T)\sum_t|\Delta[t]|$, which is the average over the absolute values of the latency at every time step.
%\begin{equation}
%\text{average latency} := \frac{1}{K}\sum^{K}_{k=1}(t_{\text{per},k}-t_{\text{est},k})\,.
%\end{equation}

\begin{figure}[t]
    \centering
    \subfigure[\emph{Normal} speed (average deviation = 83ms)]{
    \includegraphics[width=0.97\columnwidth, clip]{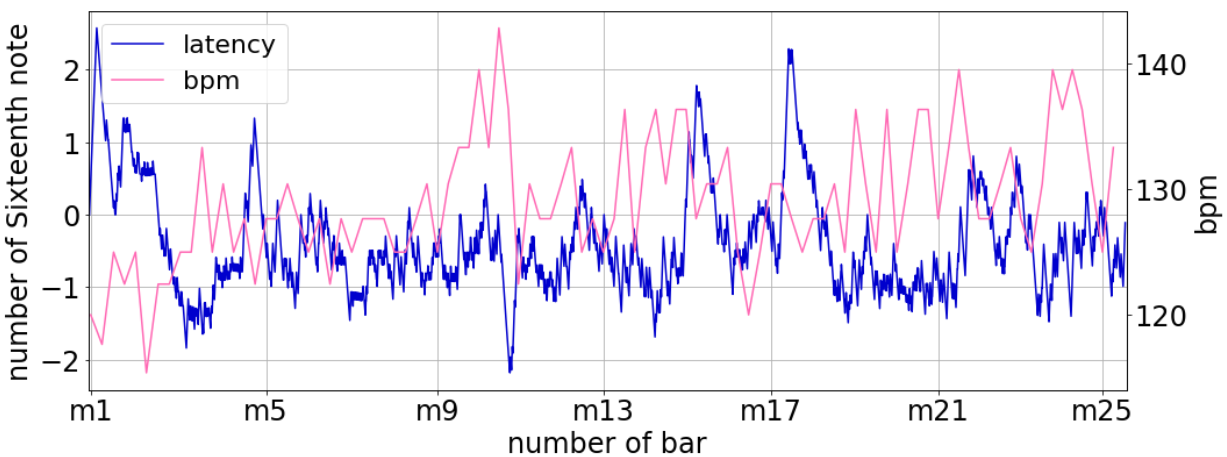}
    }
    % \vspace{-0.3cm}
    \subfigure[\emph{Slow} speed (average deviation = 145ms)]{
    \includegraphics[width=0.97\columnwidth, clip]{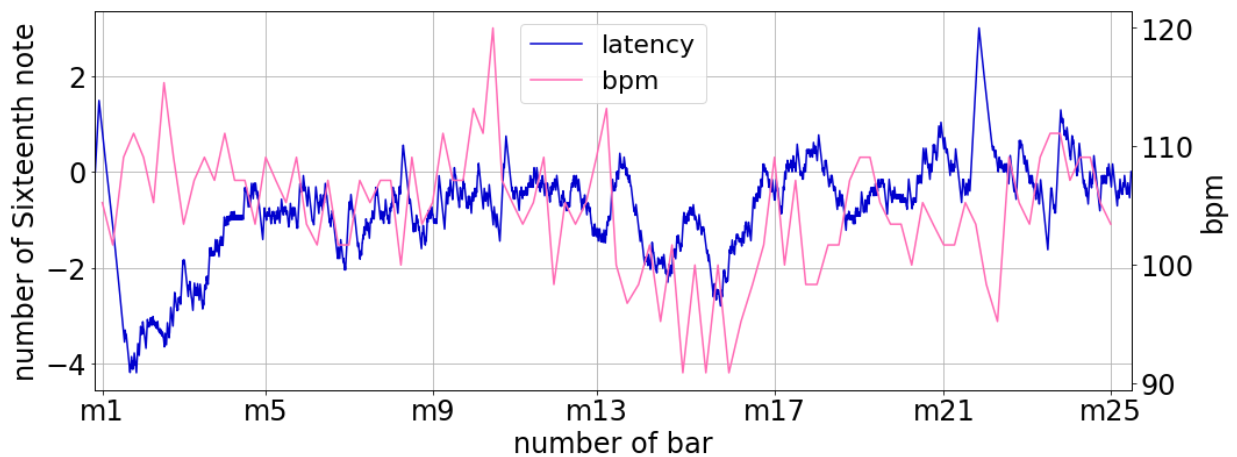}
    }
    % \vspace{-0.3cm}
    \subfigure[\emph{Fast} speed (average deviation = 124ms)]{
    \includegraphics[width=0.97\columnwidth, clip]{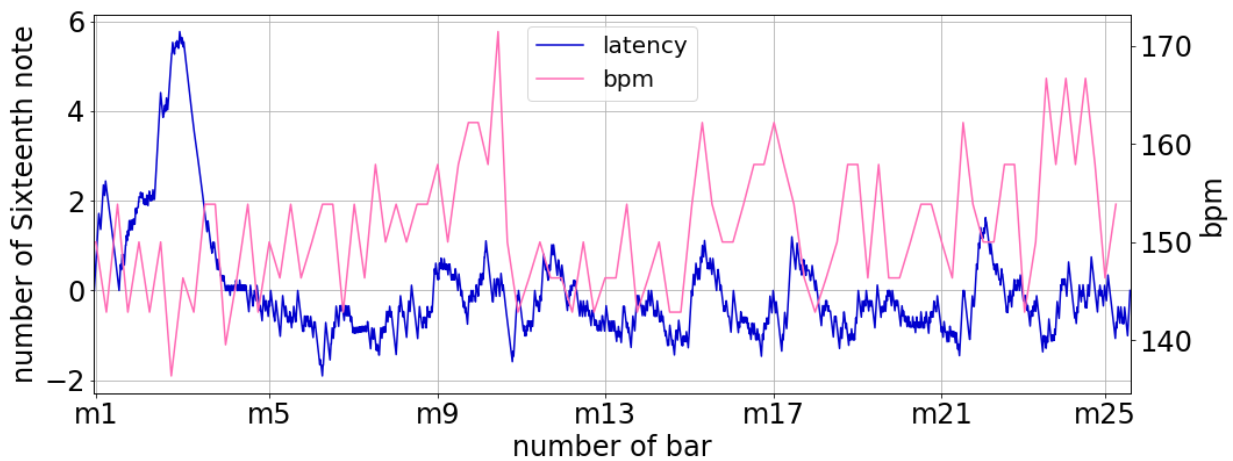}
    }
    % \vspace{-0.3cm}
    \subfigure[\emph{Accelerando} speed (average deviation = 676ms)]{
    \includegraphics[width=0.97\columnwidth, clip]{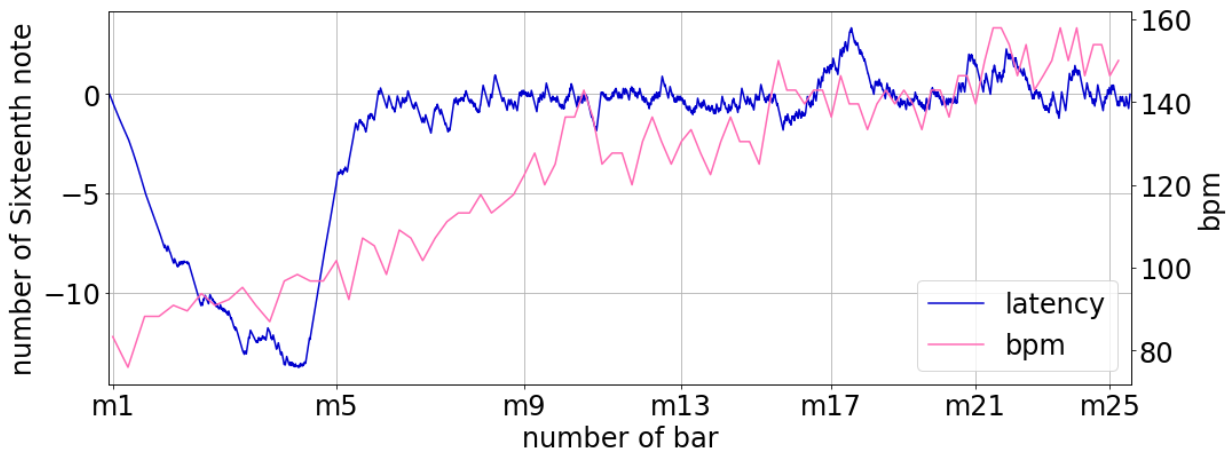}
    }
    \caption{The estimated latency with time. Blue: latency (in \#. 16th notes). Pink: tempo (in BPM).}
    \label{fig:frame_wise_latency}
\end{figure}

As shown in the upper-right part of Figure \ref{fig:recording}, %we used two distinct laptops to ensure stability during real-world performance in a concert scenario\textcolor{blue}{, as mentioned in Section 3.} 
%The real-time music tracking task is undertaken by the first laptop connecting to an audio interface, receiving the live performance piano's signal as input through a dynamic microphone. Sound generation and animation are implemented by the second laptop, which is connected to a projector to display the virtual violinist accompanying the live piano. The two laptops communicate with each other through a local network  
%In the evaluation stage, 
we used one more laptop, which records %that is in charge of recording 
both the audio streams (i.e., live piano and live violin) in the live performance. %for further analysis, connected through 
An 8-channel audio interface, Zoom UAC-8, and a robust Digital Audio Workstation (DAW), Logic Pro X, are utilized to support synchronized multi-channel recording. %The resulting audio streams are recorded synchronously into Logic Pro X, which is a robust Digital Audio Workstation (DAW) capable of accomplishing multi track recording simultaneously. 
The live piano and the reference piano MIDI are synchronized in terms of beat positions. To do this, we employ the beat tracking tool \texttt{madmom} \cite{bock2016madmom} to get the beat positions in second for both the live and reference piano. 
% \emph{performance piano} and \emph{reference piano} respectively. 
Then, the reference piano MIDI is adjusted to fit the beat positions of the live piano. %using the \texttt{adjust\_times} function in \texttt{pretty\_midi} library. 
As a result, the adjusted MIDI nicely synchronizes the live piano audio.
The estimated violin and the live violin are synchronized with conventional DTW. The \texttt{madmom} library is not used here because the beat tracking algorithm perform less accurate for violin signals.
%This adjustment yielded the estimated MIDI, which is nearly the perfect fit for the \emph{performance piano} (see Fig. \ref{fig: offline_DTW}). 
The sound of estimated violin 
is also synthesized with the timbre %from the estimated MIDI using the synthesizer 
same as with the live violin. This is to ensure stable performance of DTW synchronization with consistent audio features of both tracks. We use the chromagram features extracted from the \texttt{librosa} library for DTW synchronization. 
%Using the same tone is preferred due to using chroma feature sequence as input for DTW requires consistency in tone. %That is, our experiment showed that slight inconsistency in tone indeed yielded unexpected inaccurate alignment. 
%(Even though madmom beat tracking showed successful result in tracking piano's beat time, such tracking method failed to get the accurate beat times in the case of violin. This is the reason why we do not use madmom here.)

%\subsection{Offline DTW}
%The last step is to apply offline DTW between \emph{estimated violin} $x^{\text{vn}}_{est}$ and \emph{performance violin} $x^{\text{vn}}_{i}$, which are the two audio sequence illustrated in the bottom of Fig. \ref{fig: offline_DTW}. Our experiment showed that because violin has relatively slow attack time, using chroma feature is more suitable and accurate than the spectral flux feature, which is related to onset of the notes event.  

%We first use python library \emph{librosa} and perform STFT to both \emph{estimated violin} and \emph{performance violin}, with sampling rate of 22.05kHz, sliding window with length of 2,048 samples and hop size of 12ms. Next, we conduct offline DTW between chroma feature extracted in the previous step. %Finally, we obtain the aligned warping path of both sequences in seconds, which are two vector with same length representing each frame's time position. %By subtracting them we define as \emph{frame-wise latency}. 

%\subsection{Experiment Result}
To evaluate how the proposed system reacts under different performance speeds, we tested the system on four different ways in performing the first 25 bars in the \emph{Spring Sonata}: 1) normal speed (115-145 bpm), 2) slow speed (90-120 bpm), 3) fast speed (135-175 bpm), and 4) \emph{accelerando} speed starting with around 80bpm and ending with 160 bpm. 
Figure \ref{fig:frame_wise_latency} shows the system's latency over time by how many 16th notes the live violin delays or leads the live piano/ estimated violin. For example, in Figure \ref{fig:frame_wise_latency} (b), when the human pianist starts at a relatively slow speed of 100bpm, the live violin played faster than the benchmark (i.e. live piano) in the beginning by about four 16th notes, which is about one beat of leading. Such a situation goes stable at around the fifth measure. Similarly, in Figure \ref{fig:frame_wise_latency} (c), when the human pianist starts at a relatively fast speed of 150bpm, the live violin falls behind in the beginning, but also get synchronized after the fifth measure, where the latency values within one 16th note can be observed. In summary, for all the four cases, it takes the system around four measures to get synchronized with the live piano, and once it gets synchronized with the human musician, the deviation can be found within $\pm$0.25 beats. We can also observe that an abrupt change of speed does not always imply a change of latency. In fact, latency is also related to the structure of music; for example, the latency values for the four cases are relatively unstable at around the 15th measure, where the piano part is the main melody in a rapid note sequence. These unstable parts increase the average deviation for each music sample and cause few but clear mismatch, as demonstrated in the supplementary video.

\begin{table}[t]
\caption{The results of body movement generation.}
\begin{tabular}{|c|cccc|}
\hline
      & $x$-axis   & $y$-axis   & $z$-axis & average \\ \hline\hline
model & 0.3531 & 0.4132 & 0.3300 & 0.3655   \\
beat & 0.1666 & 0.1655 & 0.1684 & 0.1669 \\
\hline
\end{tabular}
\label{tab:body_results}
\vspace{-0.5cm}
\end{table}
% L1 = 0.0393, elbow length = 0.13

\subsection{Assessment of body movement generation}

%\subsection{Dataset}
The model is trained and validated on a recorded dataset containing 140 %videos 420 videos (about 11 hours)\textcolor{red}{(how many hours?)} 
violin solo videos with total length of 11 hours. These videos are from 10 college students who major in violin. 14 violin solo pieces was selected to be performed by the 10 students; that means, each violin solo piece has 10 versions in the dataset. %and had learned violin more than 10 years. %with Taipei National University of the Arts (TNUA). In this dataset, we recruited 10 participants who had experience in playing violin more than 10 years and gave them 14 musical pieces to play. All participants had fully practiced each musical piece before shooting videos. 
%Each performance was shot by three 1080p cameras from different directions: front, left and right side, and the FPS is set to 30. In this paper, we only use front side of the performances. 
The details of this dataset will be announced aftereards. %this paper is published.

For training, we first select the videos from one of the violinists, and conduct a leave-one-piece-out (i.e., 14-fold cross validation) scheme to train the 14 models. Each model is evaluated on the left-out piece (for that fold partition) performed by the remaining nine violinists. The results are then the average over the nine violinists. That means, our evaluation strategy benefits from multiple versions of each music piece in the dataset.

%For body movement generation, we use leave-one-piece-out scheme to conduct 14-fold cross validation with 14 musical pieces and choose the performances played by one of the violinists (No.4 participant) who have stable performance in average to be training data and the others to be test data. All of the quantitative evaluations are calculated by averaging the results of 14 fold and 9 violinists, as shown in 

Two metrics, $L_1$ distance and bowing attack F1-score, are used for quality assessment. $L_1$ distance is simply the distance between the generated and the ground-truth joints. 
%, and bowing attack accuracy is to evaluate the rationality. 
In the practice of music performance, a bowing attack is the time instance when the bowing direction changes (i.e., from up-bow to down-bow or from down-bow to up-bow). In other words, bowing attack is an important feature regarding how reasonable the generated body movement is. The bowing attack of the ground truth and the generated results can therefore be estimated by the first-order difference of the right-hand wrist joint sequence. By abuse of this definition, we define a bowing attack in the $x$, $y$, or $z$ direction as the time instance that the right-hand wrist joint changes its direction in the $x$, $y$, or $z$ direction, respectively. For a ground-truth bowing attack at $i$, a predicted bowing attack is a true positive if it is in the interval of $[i-\delta, i+\delta]$. If there are two or more predicted bowing attacks in that interval, only one of them is regarded as true positive and others are regarded as false positive. After identifying all the true positive, false positive, and false negative, the F1-scores for the bowing attack in the three directions can be reported. In this paper, we set $\delta=1$ frame, i.e., 1/30 seconds.

%of generated results and ground truth as $\hat{y}^{(rh)}$ and $y^{(rh)}$, where $\hat{y}^{(rh)}$, ${y^{(rh)}}$ $\in \mathbb{R}^3$. We then calculate the direction $D(i)$ for both sequence as:

%\begin{equation}\label{eq:9}
%D(i) =
%  \begin{cases}
%    1  & \quad \text{if } y^{(rh)}(i+1) - y^{(rh)}(i) > 0,\\
%    0  & \quad \text{if } y^{(rh)}(i+1) - y^{(rh)}(i) \leq 0.
%  \end{cases}
%\end{equation}

%Accordingly, we separately get the direction of right-hand wrist joint for generated results $\hat{D}(i)$ and ground truth $D(i)$. Derived from the bowing direction $D(i)$, the bowing attack $A(i)$ at time i would be set 1 if the direction $D(i)$ is different from $D(i-1)$:
%\begin{equation}
%\label{eq:10}
%A(i) =
%  \begin{cases}
%    1  & \quad \text{if } D(i) - D(i-1) \neq 0,\\
%    0  & \quad \text{otherwise }.
%  \end{cases}
%\end{equation}

%Finally we get bowing attack $\hat{A}(i)$ and $A(i)$. we here claim that the predicted bowing attack $\hat{A}(i)$ is correct if real bowing attack is located on the range $[i-\delta, i+\delta]$. Otherwise, it would be failed if there is no any real bowing attack in range $[i-\delta, i+\delta]$. Notice that all real bowing attacks are only calculated once. If all real bowing attacks near $\hat{A}(i)$ have been calculated before, we will claim $\hat{A}(i)$ is failed. Here $\delta$ is set to 1. After computing bowing attack label, we use them to calculate F1-score.

The average $L_1$ distance is 0.0393; to be more specific, it is around 30\% of the lower arm length (average value = 0.13). %This is small
Table~\ref{tab:body_results} lists the resulting F1 scores of bowing attack and the average over the three directions, compared to the baseline results, which take naive guess by taking all beat positions as the bowing attack (denoted as `beat' in Table~\ref{tab:body_results}). We observe that the F1-scores are much better than the baseline, which shows the effectiveness of the model. The F1 score in the $y$-direction performs better than others, since the actual bowing direction is mainly in the $y$-axis (the front-back direction of the violinist). To understand the actual quality of these results more, in the supplementary material we compare two animation results, one using the skeleton of pose estimation from a rehearsed video, and the other using the skeleton generated from the body movement generation model. It shows that the deviation of the right hand position and the timing of the bowing attack are the two issues that still have room for improvement, which are consistent with the qunatitative experiment results.

\vspace{-0.3cm}

%From the video recordings, we show two versions of animation results, one using pose estimation from a rehearsed violin video, and the other using the results of the body movement generation model. 
%\subsection{Model description}
%\textbf{Training procedure}
%We take 13 musical pieces from one of the violinists as the training data and the remaining one being the validation data to conduct 14-fold cross validation. 
%\input{Paragraph/Discussion}
\section{Conclusion} \label{sec:conclusion}

A virtual musician system supporting human-computer duet from audio has been implemented. We have accomplished fully automatic real-time tracking and virtual performance, both controlled by music audio only. By integrating all the novel techniques together, we then reexamine the limitation of these techniques at a higher level, such as the mismatch of the duet performance at the beginning of the performance, and the instability of the end-to-end body movement generation method, both of which cannot be observed with simplified evaluation metrics. These provide new insights and possible ways to improve the system, such as incorporating an online feedback mechanism to suppress abrupt change of tracking position, and combining end-to-end rigging and motion generation model such as rignet \cite{xu2020rignet} for more precise animation. Demo videos, supplementary materials, and acknowledgement can be seen at the project website: \url{https://sites.google.com/view/mctl/research/automatic-music-concert-animation}

%will be the future direction

%New insights and issues are also revealed from the integrated system and are found from the quality assessment stage,.
%Since we challenge the tasks to use audio rather than MIDI signal for real-time tracking, such a system is applicable for general kinds of instruments other than piano. 

%1) The cold-start problem of tracking, 2) The motion generation is still in development stage, 


%% \bibliographystyle{ACM-Reference-Format}
\bibliographystyle{unsrt}
\bibliography{references}

%\appendix

\end{document}